\documentclass{egpubl}
\usepackage{eg2017}

  \ShortPresentation      
  \electronicVersion 

\ifpdf \usepackage[pdftex]{graphicx} \pdfcompresslevel=9
\else \usepackage[dvips]{graphicx} \fi

\PrintedOrElectronic

\usepackage{t1enc,dfadobe}

\usepackage{egweblnk}
\usepackage{cite}

\usepackage{amsmath}
\usepackage{algorithm}
\usepackage[noend]{algpseudocode}
\usepackage[inline]{enumitem}
\usepackage{verbatim}
\usepackage{gensymb}

\title[AIR: Anywhere Immersive Reality with User-Perspective Projection]%
      {AIR: Anywhere Immersive Reality with User-Perspective Projection}

\author[JH. Byun, SH. Chae, YS. Yang \& TD. Han]
{\parbox{\textwidth}
	{\centering 
		JungHyun Byun, SeungHo Chae, YoonSik Yang and TackDon Han
	}
        \\
{\parbox{\textwidth}
	{\centering 
		Media System Lab., Department of Computer Science, Yonsei University, Korea, Republic of
       } 
}
}

\begin{document}
\maketitle

\begin{abstract}
	Projection-based augmented reality (AR) has much potential, but is limited in that it requires burdensome installations and prone to geometric distortions on display surface.
	To overcome these limitations, we propose \textbf{AIR}. 
	It can be carried and placed anywhere to project AR using pan/tilting motors, while providing the user with distortion-free projection of a correct 3D view.


\end{abstract}

\section{Introduction}

Many augmented reality (AR) applications are built on head-worn or hand-held see-through displays, accompanied by issues of the limited field of view (FOV) and the user fatigue from bulky equipments\cite{bimber2005spatial}.
Projection-based AR in contrast utilizes a projector to solve these issues. Located off-site, a projector directly augments graphics of virtually unlimited sizes and FOV onto physical objects \cite{cebulla2013projection} immersing users' surroundings, without diverting their attention from the real world\cite{benko2014dyadic}. However, these features come at a price \cite{bimber2005embedded}. First, direct augmentation with projection is prone to geometric distortion. Second, a detached projector renders graphics that are inconsistent with the user's viewpoint.

The most crucial problem with projection AR is that the overall quality of AR relies heavily on the characteristics of the surface\cite{raskar1998spatially}. 
A diffuse object with smooth geometry is required for an ideal projection, but that is rarely the case in projection AR\cite{raskar1998office}. 
Thus, the projection area is constrained to the size and shape of the objects' surfaces\cite{bimber2005spatial}. Projection on unideal surfaces with bumps and gaps would otherwise yield distortions.

Another problem in projection AR as well as most AR systems is inconsistency between the real scene and rendered graphics. Graphics are rendered on captured images from the perspective of the device, not the user. 
The former is called device-perspective rendering (DPR), and the latter user-perspective rendering (UPR)\cite{tomioka2013approximated}. DPR diminishes the sense of presence of virtual contents and this sense diminishes further as the camera-eye discrepancy, in terms of FOV, distortion, and viewpoint, increases. 
Because in projection AR, a projector is located independently of the user\cite{bimber2005spatial}, the risk of diminished realism is much more severe.

To address these issues and realize immersive projection AR, we propose 
\emph{Anywhere Immersive Reality (AIR)}, a novel projection system for any environment with user perspective rendering.
The proposed system is a steerable platform with two cameras and a projector. It is designed to be carried and placed anywhere, rotated in any direction, and project contents even on uneven surfaces, while providing a correct 3D perspective view without distortions.

\section{Related Work}

\subsection{Immersive Projection Display}

In spatially immersive display (SID), users are surrounded by planar or curved screens, onto which multiple projectors display. Studies like \cite{cruz1993surround}, or \cite{benko2012miragetable} implemented stationary versions of SID which require users to be at predesignated point for immersion.
\cite{benko2014dyadic} and \cite{jones2014roomalive} extended this approach to a room-scale, non-planar geometry. Users can move around the room and be provided with correct 3D virtual views regardless of the surface geometry. However, since these SID approaches require extensive workloads of physical installations, off-line calibrations, and many high-cost hardware, they are not appropriate for the everyday user.

In this study, we also aim to realize an immersive projection system, but on an accessible budget. Our system is a pan-tilting platform, on which a projector-camera system for AR is built, with an additional camera unit for user tracking. Rotating the platform makes it possible to project virtually anywhere without the necessity of multiple units of costly hardware. 

\subsection{Portable Projection System}

Most works on projection AR have predominantly focused on infrastructure-based spatial-aware systems, which infer the location and orientation of the device in 3D space based on the notion of one or more environmental sensors \cite{raskar2006ilamps}. Recently, the miniaturization of pico-projectors made it possible for systems to be portable, serving as an enabling technology for ubiquitous computing \cite{huber2014research}. For a system to be truly portable, it should be not only spatially aware, but also geometrically aware, i.e., the system constructs the 3D structure of the world, and self-contained \cite{molyneaux2012interactive}. 

We also aim to design such a portable projection system. However, most of these systems are handheld, which can cause fatigue and limitation in interactions as users have to hold on to the system at all times. To solve these problems, the proposed system can be carried and placed, or held, and provide a seamless display.

\subsection{Projection Distortion Correction}

Many approaches have been proposed to overcome the effect of surface geometry on projection.
In \cite{raskar2001self} and \cite{xiao2013worldkit}, systems pre-warp images using the homography between a projector and its display surface, so that the projected image appears as it is desired. 
These systems, however, are limited as it can only correct oblique images on planar surfaces from a certain fixed device viewpoint. 

Both \cite{pinhanez2001everywhere} and \cite{wilson2012steerable} utilize pan-tilting platforms for a rotatable viewpoint, but only the latter can adapt to dynamic environments and users, and correct complex distortions accordingly. 
However, pre-construction of the room geometry is required for the system, and localization of the user via sound demands even further installations and calibrations. 
Beyond these limitations, our system uses only real-time, raw depth data, and still project immersive, distortion-free AR content without prior knowledge of the space.

The main contributions of this paper are summarized as follows:
\begin{itemize} 
\item We propose the mathematical modeling of such a system, from calibration to implementation.
\item We propose the concept of a portable projection system, and its possible applications.
\end{itemize}

\section{Proposed System}

\begin{figure}[htb] 
  \centering
  \includegraphics[width=.96\linewidth]{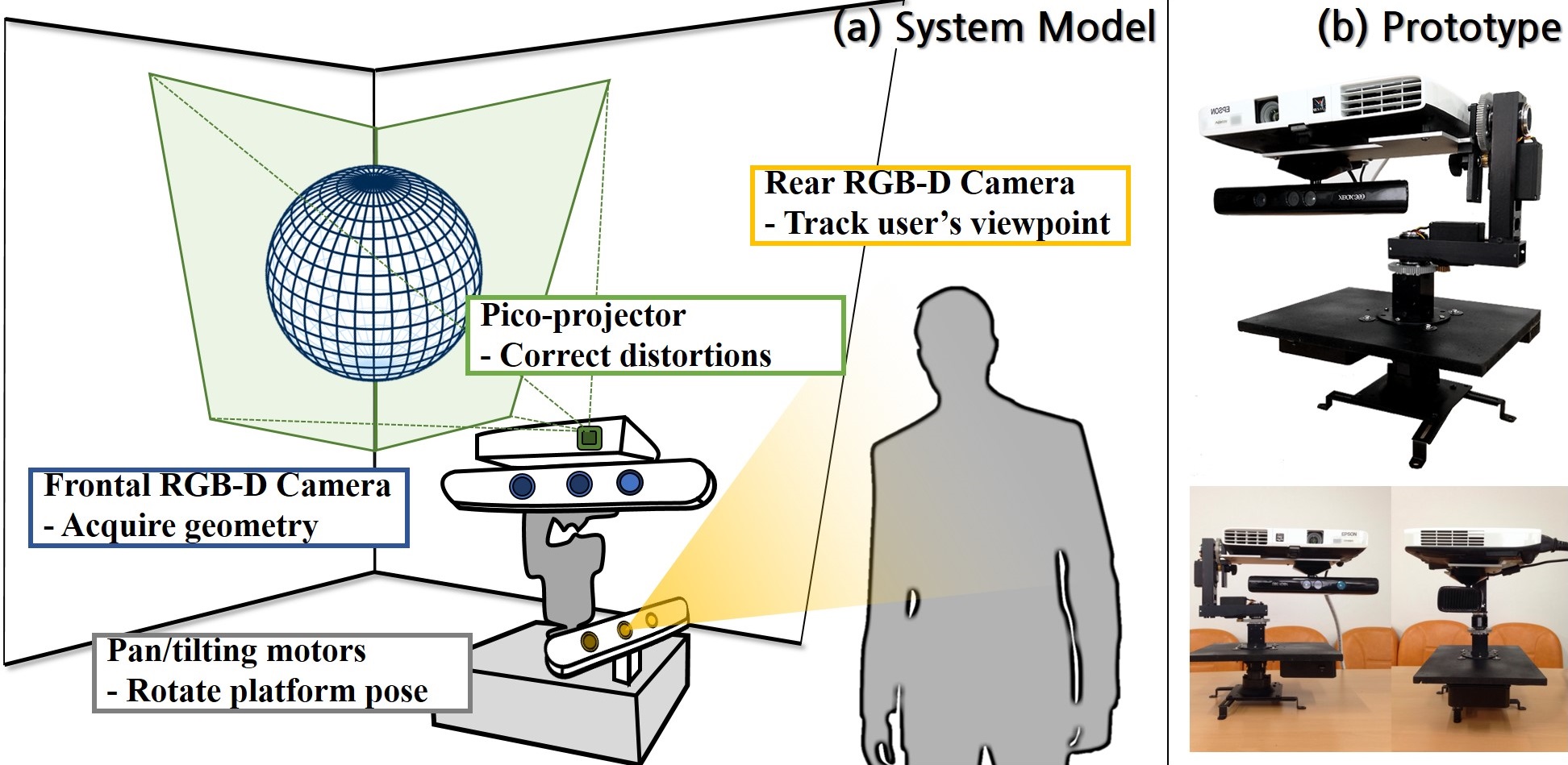}
  \caption{\label{fig:system_model}
          Proposed \textbf{(a)} system model and \textbf{(b)} its prototype}
\end{figure}

\subsection{System Overview}

Figure~\ref{fig:system_model} describes the system model of \emph{AIR}. The proposed system is based on a projector-camera system. It is held up by a platform, to which servo motors are attached vertically and horizontally, so that it can pan or tilt freely. We added a rear-facing RGB-D camera that serves as a dedicated user tracker for user-perspective rendering.

We built a prototype on a personal computer with an Intel i7-3770 CPU, 8GB RAM, and an NVIDIA GTX 960 GPU. The system consists of an Epson projector, a  Microsoft Kinect v2, a Microsoft Kinect 360, servo motors, and an Arduino board for control.

\subsection{System Flow}

\begin{figure}[htb]
  \centering
  \includegraphics[width=0.99\linewidth]{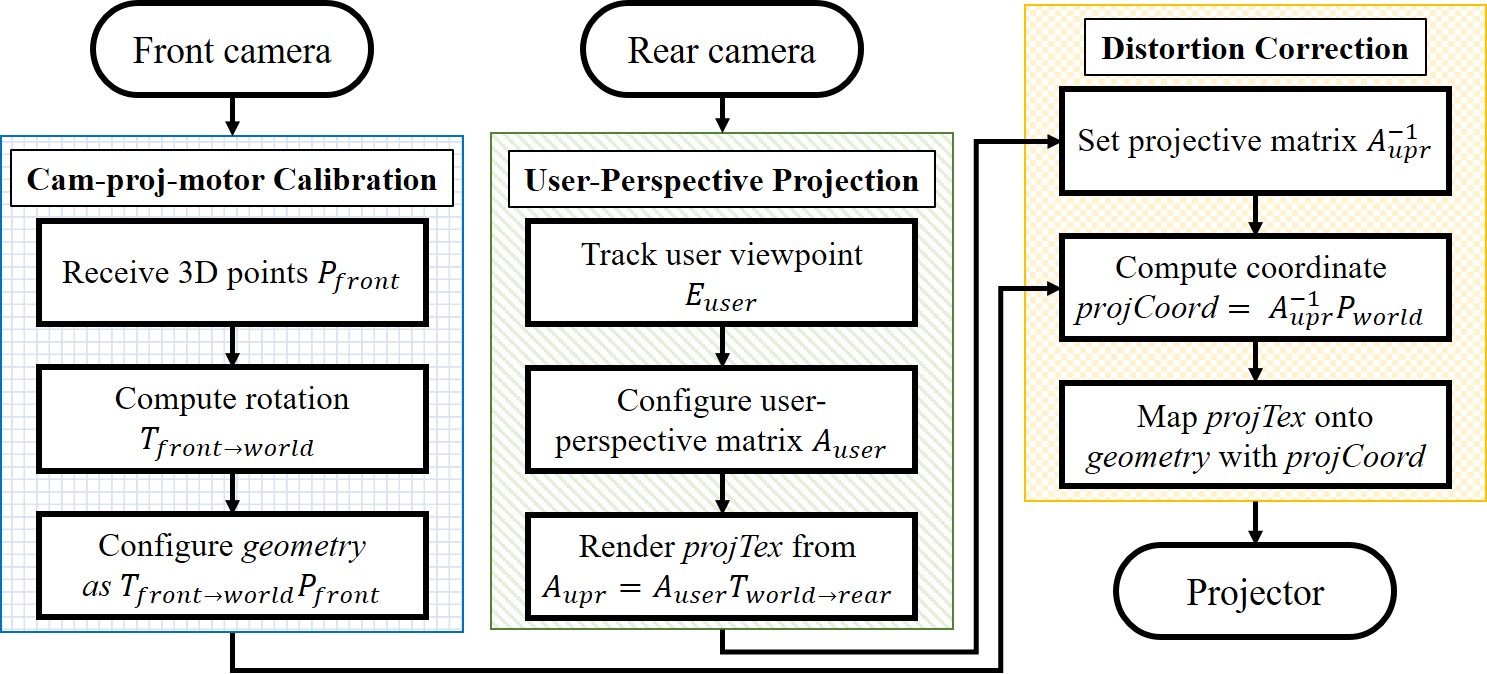}
  \caption{\label{fig:overall_process}
          System workflow}
\end{figure}
In this section, we present the mathematical model of \emph{AIR}, from calibrations to implementations. We extend the algorithm for a projector-camera system in \cite{raskar1998office} to incorporate rotatable motors and an additional camera, and handle dynamic environments and users.
The three core steps to realize user-perspective projection are 
\begin{enumerate*}[label=\roman*)]
\item camera-projector-motor calibration,
\item user-perspective projection, and
\item distortion correction.
\end{enumerate*}

\subsubsection{Camera-projector-motor Calibration}

\begin{figure*}[!h]
 \includegraphics[width=0.99\linewidth]{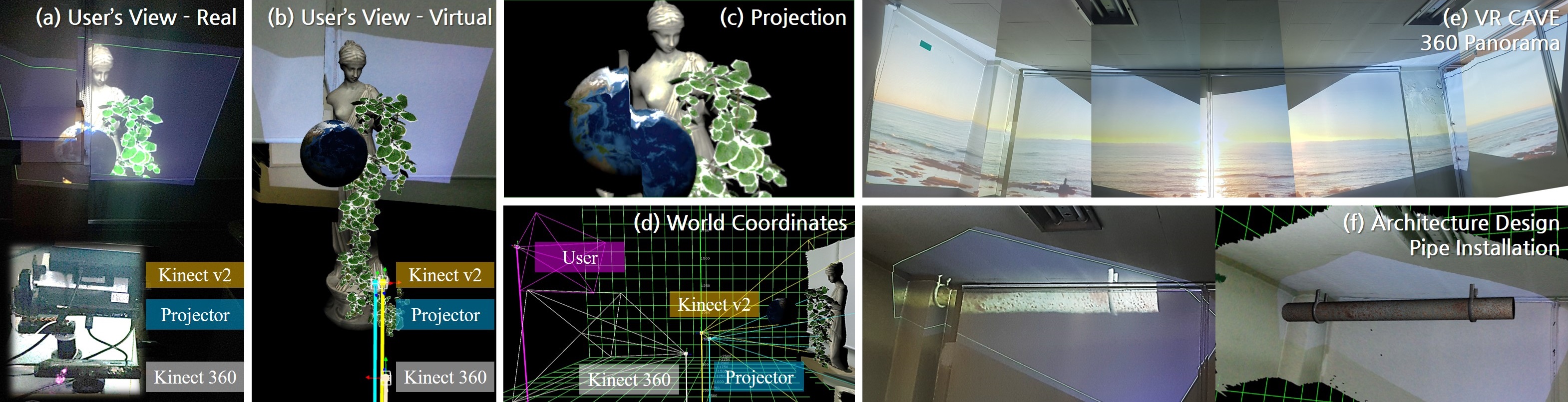}
  \caption{ 
	\textbf{(a)} \textbf{AIR} provides the user with correct projection of \textbf{(b)} a 3D perspective view.
	\textbf{(c)} The projection is pre-warped to compensate for distortions.
	\textbf{(d)} Cameras and a projector are registered in world. 
	Applications for \textbf{AIR} include \textbf{(e)} VR Cave and \textbf{(f)} architecture design.
  }
\label{fig:teaser}
\end{figure*}
\emph{AIR} consists of two cameras, two motors, and one projector, each of which have their own coordinate systems. The first step is to configure them in one common $world$ coordinate system. We setup the origin of the $world$ system as the pose of front-facing camera facing front, with no rotation.
For simplicity, we assume that internal parameters $A_{front}$ and $A_{rear}$ of the front and rear camera are known.

First, we model the movements of the pan/tilting motors. Because their rotational axes are not exactly perpendicular and parallel to the ground, we calculate the rotation matrix $R(\theta)$ of angle $\theta$ with an arbitrary, unit axis vector $(x,y,z)$ as\\
\resizebox{\hsize}{!}{$
\begin{bmatrix}
\cos\theta+x^2(1-\cos\theta) & xy(1-\cos\theta)-z\sin\theta & xz(1-\cos\theta          )+y\sin\theta \\
yx(1-\cos\theta)+z\sin\theta & \cos\theta+y^2(1-\cos\theta) & yz(1-\cos\theta          )-x\sin\theta \\
zx(1-\cos\theta)-y\sin\theta & zy(1-\cos\theta)+x\sin\theta & \cos\theta+z^2(1-\cos\theta)
\end{bmatrix}.
$}
A 3D point $P_{world}$ in $world$ coordinates and its corresponding $P_{front}$ perceived by the front camera can be modeled as 
\begin{equation} \label{eq:front_to_world}
P_{world} = R(\beta)R(\alpha)P_{front},
\end{equation}
 with panning angle $\alpha$ and tilting $\beta$. Matrix product $R(\beta)R(\alpha)$ can be thought of as external parameters of the front camera, denoted as $T_{front\to\,world}$.  We use a checkerboard of a known size at a fixed pose and solve for $x,y,z$ with least mean square method for two rotational axes while rotating the system about each of those axes.

The next step is the calibration of the rear-facing camera. The projector-camera system is rotated to a certain angle so that there is enough overlap between the front camera and rear camera for calibration. 
Using the coordinates of the checkerboard in equation~\ref{eq:front_to_world} as the baseline, the relations among the $world$, front camera, and rear camera coordinate systems can be expressed as
\begin{equation} \label{eq:rear_to_world}
P_{world} = T_{rear\to\,world}P_{rear} = R(\beta)R(\alpha)P_{front}.
\end{equation}

Finally, the calibration between the front-facing RGB-D camera and the projector is performed. We used the calibration method proposed in \cite{kimura2007projector}. A projector can be thought of as a reverse-camera model with the same parameters. The projector projects chessboards, $x_{proj}$, of various, pre-known sizes to different planar surfaces. Applying standard projective geometry, we obtain
\begin{equation} \label{eq:front_to_proj}
sx_{proj} = A_{proj}T_{front\to\,proj}P_{front}.
\end{equation}
Since depth data $P_{front}$ are available, we can solve for the internal parameters of projector $A_{proj}$ and external parameters $T_{front\to\,proj}$ of the rotations and translations with regard to the front camera.

\subsubsection{User-Perspective Projection}

Providing the user with a correct 3D perspective view and motion parallax is the key factor in realizing immersive AR. We setup the equation of user-perspective rendering\cite{tomioka2013approximated}. Instead of an actual tablet, here the screen is located in the virtual world. The system tracks poses of the user's head from the rear camera. The user's viewpoint is denoted as $E_{user} = [e_x, e_y, e_z]^T$, and its perspective projection matrix $A_{user}$. The corresponding equation is as follows:
\begin{equation} \label{eq:user_to_world}
A_{upr} = A_{user}T_{world\to\,rear}, \text{where } A_{user} = 
\setlength\arraycolsep{1pt}
\begin{bmatrix}
-e_{z} &0 &e_{x} &0 \\ 
0 &-e_{z} &e_{y} &0 \\ 
0 &0 &1 &-e_{z}    
\end{bmatrix}.
\end{equation}
 
\subsubsection{Distortion Correction}

Unlike mobile AR that uses a tablet screen, we use an off-site projector. This causes discrepancies in viewpoints, projection matrices and so on, which result in distorted projection. To correct these distortions, we implemented a two-pass rendering pipeline for projective texture mapping\cite{everitt2001projective}, where we simulate the modeling of a calibrated projector-camera system. First, the system receives the geometry of the real world, and virtual objects are rendered from the user's perspective. Then, the rendered result is used as the texture to be mapped projectively onto the geometry. The final rendering is correct in terms of both user-perspective and geometric distortions. Figure~\ref{fig:overall_process} shows the overall process.

\subsection{Applications}

In this section, we describe some possible applications for \emph{AIR}. Projection-based AR has a lot of potentials, as it can directly augment physical objects with graphics, promoting interactive cooperation of multiple users.
Figure~\ref{fig:teaser} shows our early implementation of \emph{AIR} including 360 panorama and pipe installation applications. 

Previously to be immersed in 360 images, systems like VR-CAVE\cite{cruz1993surround} were needed, which require large spaces, screens, and multiple projectors, or bulky head mounted displays (HMD), which restrict shared experiences.
To ease these limitations, we developed a 360 panorama viewer (Figure~\ref{fig:teaser}(e)) on \emph{AIR} with minimum hardware. It can project correct rotating views of a 360 image on surfaces and users can enjoy them together.

Another field in which \emph{AIR} can be utilized is architecture design (Figure~\ref{fig:teaser}(f)). 
Previous HMD-based systems required off-line reconstruction of the room geometry and forced solo experience. 
In contrast, \emph{AIR} system can be placed on site and project actual-scale virtual objects, such as a pipe. Clients and designers together can get a good sense of how it will turn out before installation.

\section{Evaluation and Results}

In \cite{benko2012miragetable} to evaluate the effect of distortion correction, authors computed root mean square errors of absolute differences of corrected projections pixel-by-pixel. However, as stated in the paper, directly measuring pixel differences is susceptible to varying color conditions, while our only concern here is geometric distortion.

To measure errors without the influence of color conditions, we compute differences in structural compositions, point-by-point. First, we setup the base image by projecting a checkerboard on a planar surface. Then, we placed different obstacles before the surface and computed the correct projections. We matched corresponding corner points $P_i$ of the checkerboards with those of the base image and computed the dislocations of the pixels using
\begin{equation}
\frac{1}{N}\displaystyle\sum_{i=1}^{N} |P_{base}-P_i|
\text{, where $N$: the number of corners}.
\end{equation}
\begin{figure}[htb]
  \centering
  \includegraphics[width=.99\linewidth]{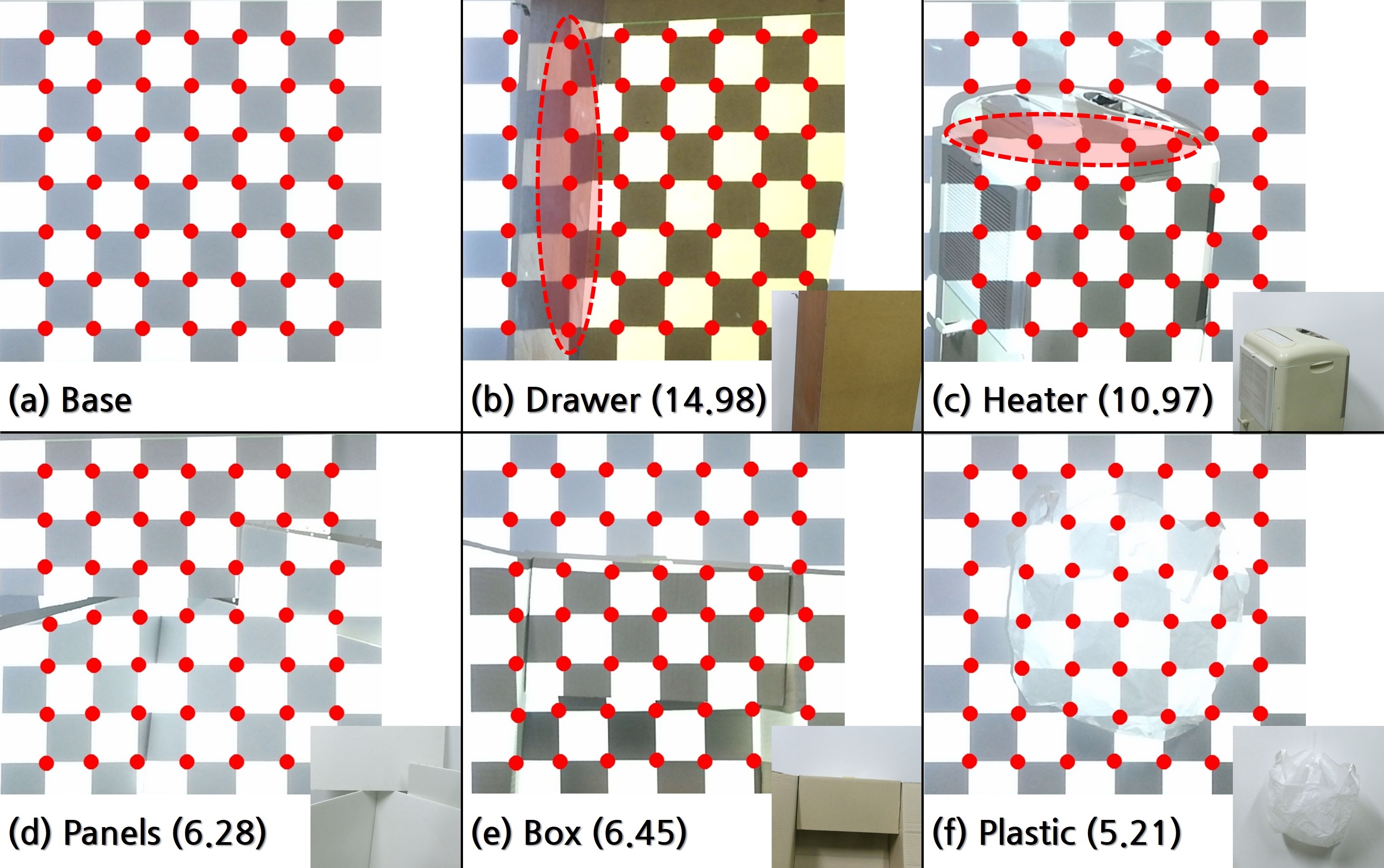}
  \caption{\label{fig:eval_results}
          Evaluation results of five obstacles with the base image. Corner points of the checkerboards are denoted in red dots.}
\end{figure}
Figure~\ref{fig:eval_results} shows results of the evaluation. Only projection regions of original 1920 by 1080 images are cropped. Numbers on the caption are average dislocation errors in pixel. 

We tested our method on obstacles of various geometric attributes. 
Results show that our method corrects distortions in projection effectively even on (d), (e) complex geometry and (f) a freely deformable object. Rigid (b) and curved geometry (c) caused the highest errors. 
This is an expected result, as objects with a flat angle to the camera lead to a complete lack of depth information. Specifically, for angles between 10\degree and 30\degree the Kinect delivers up to 100\% invalid pixel\cite{sarbolandi2015kinect}. As one can see in (b) and (c), dislocations were greatest on slopes that are almost parallel to the depth camera (red dotted region), resulting in wrong depth value and wrong distortion correction accordingly. Since this is innate limitation of the Kinect, we leave this problem as our future work.

\section{Conclusion and Future Work}

We presented \emph{AIR}, a steerable projection system for anywhere immersive reality. Our system can be carried, placed, and project anywhere without prior knowledge of the room geometry, and still deliver a correct 3D view of the virtual world to the user. We also implemented two applications in which the \emph{AIR} system would be useful. An evaluation on projection distortion correction was conducted and the results confirm the usability of \emph{AIR}.

We intend to further the research on \emph{AIR} by integrating the system with natural user interactions so that users can be immersed in and interact with an AR space naturally and intuitively. We also intend to broaden and validate our search for application scenarios.

\section{Acknowledgments}

This work was supported by the National Research Foundation of Korea(NRF) grant funded by the Korea government(MSIP) (No. NRF-2015R1A2A1A10055673).

\bibliographystyle{eg-alpha-doi}

\bibliography{sources/egbibsample}

\newcommand{\etalchar}[1]{$^{#1}$}
\begin{thebibliography}{\uppercase{RVBB{\etalchar{*}}06}}

\bibitem[BEK05]{bimber2005embedded}
\textsc{Bimber O., Emmerling A., Klemmer T.}:
\newblock Embedded entertainment with smart projectors.
\newblock \emph{Computer 38}, 1 (2005), 48--55.

\bibitem[BJW12]{benko2012miragetable}
\textsc{Benko H., Jota R., Wilson A.}:
\newblock Miragetable: freehand interaction on a projected augmented reality
  tabletop.
\newblock In \emph{Proceedings of the SIGCHI conference on human factors in
  computing systems} (2012), ACM, pp.~199--208.

\bibitem[BR05]{bimber2005spatial}
\textsc{Bimber O., Raskar R.}:
\newblock \emph{Spatial augmented reality: merging real and virtual worlds}.
\newblock CRC press, 2005.

\bibitem[BWZ14]{benko2014dyadic}
\textsc{Benko H., Wilson A.~D., Zannier F.}:
\newblock Dyadic projected spatial augmented reality.
\newblock In \emph{Proceedings of the 27th annual ACM symposium on User
  interface software and technology} (2014), ACM, pp.~645--655.

\bibitem[Ceb13]{cebulla2013projection}
\textsc{Cebulla A.}:
\newblock Projection-based augmented reality.
\newblock \emph{ETH Zurich} (2013).

\bibitem[CNSD93]{cruz1993surround}
\textsc{Cruz-Neira C., Sandin D.~J., DeFanti T.~A.}:
\newblock Surround-screen projection-based virtual reality: the design and
  implementation of the cave.
\newblock In \emph{Proceedings of the 20th annual conference on Computer
  graphics and interactive techniques} (1993), ACM, pp.~135--142.

\bibitem[Eve01]{everitt2001projective}
\textsc{Everitt C.}:
\newblock Projective texture mapping.
\newblock \emph{White paper, NVidia Corporation 4} (2001).

\bibitem[Hub14]{huber2014research}
\textsc{Huber J.}:
\newblock A research overview of mobile projected user interfaces.
\newblock \emph{Informatik-Spektrum 37}, 5 (2014), 464--473.

\bibitem[JSM{\etalchar{*}}14]{jones2014roomalive}
\textsc{Jones B., Sodhi R., Murdock M., Mehra R., Benko H., Wilson A., Ofek E.,
  MacIntyre B., Raghuvanshi N., Shapira L.}:
\newblock Roomalive: magical experiences enabled by scalable, adaptive
  projector-camera units.
\newblock In \emph{Proceedings of the 27th annual ACM symposium on User
  interface software and technology} (2014), ACM, pp.~637--644.

\bibitem[KMK07]{kimura2007projector}
\textsc{Kimura M., Mochimaru M., Kanade T.}:
\newblock Projector calibration using arbitrary planes and calibrated camera.
\newblock In \emph{2007 IEEE Conference on Computer Vision and Pattern
  Recognition} (2007), IEEE, pp.~1--2.

\bibitem[MIK{\etalchar{*}}12]{molyneaux2012interactive}
\textsc{Molyneaux D., Izadi S., Kim D., Hilliges O., Hodges S., Cao X., Butler
  A., Gellersen H.}:
\newblock Interactive environment-aware handheld projectors for pervasive
  computing spaces.
\newblock In \emph{International Conference on Pervasive Computing} (2012),
  Springer, pp.~197--215.

\bibitem[Pin01]{pinhanez2001everywhere}
\textsc{Pinhanez C.}:
\newblock The everywhere displays projector: A device to create ubiquitous
  graphical interfaces.
\newblock In \emph{International Conference on Ubiquitous Computing} (2001),
  Springer, pp.~315--331.

\bibitem[RB01]{raskar2001self}
\textsc{Raskar R., Beardsley P.}:
\newblock A self-correcting projector.
\newblock In \emph{Computer Vision and Pattern Recognition, 2001. CVPR 2001.
  Proceedings of the 2001 IEEE Computer Society Conference on} (2001), vol.~2,
  IEEE, pp.~II--504.

\bibitem[RVBB{\etalchar{*}}06]{raskar2006ilamps}
\textsc{Raskar R., Van~Baar J., Beardsley P., Willwacher T., Rao S., Forlines
  C.}:
\newblock ilamps: geometrically aware and self-configuring projectors.
\newblock In \emph{ACM SIGGRAPH 2006 Courses} (2006), ACM, p.~7.

\bibitem[RWC{\etalchar{*}}98]{raskar1998office}
\textsc{Raskar R., Welch G., Cutts M., Lake A., Stesin L., Fuchs H.}:
\newblock The office of the future: A unified approach to image-based modeling
  and spatially immersive displays.
\newblock In \emph{Proceedings of the 25th annual conference on Computer
  graphics and interactive techniques} (1998), ACM, pp.~179--188.

\bibitem[RWF98]{raskar1998spatially}
\textsc{Raskar R., Welch G., Fuchs H.}:
\newblock Spatially augmented reality.
\newblock In \emph{First IEEE Workshop on Augmented Reality (IWAR 98)} (1998),
  Citeseer, pp.~11--20.

\bibitem[SLK15]{sarbolandi2015kinect}
\textsc{Sarbolandi H., Lefloch D., Kolb A.}:
\newblock Kinect range sensing: Structured-light versus time-of-flight kinect.
\newblock \emph{Computer Vision and Image Understanding 139} (2015), 1--20.

\bibitem[TIS13]{tomioka2013approximated}
\textsc{Tomioka M., Ikeda S., Sato K.}:
\newblock Approximated user-perspective rendering in tablet-based augmented
  reality.
\newblock In \emph{Mixed and Augmented Reality (ISMAR), 2013 IEEE International
  Symposium on} (2013), IEEE, pp.~21--28.

\bibitem[WBIH12]{wilson2012steerable}
\textsc{Wilson A., Benko H., Izadi S., Hilliges O.}:
\newblock Steerable augmented reality with the beamatron.
\newblock In \emph{Proceedings of the 25th annual ACM symposium on User
  interface software and technology} (2012), ACM, pp.~413--422.

\bibitem[XHH13]{xiao2013worldkit}
\textsc{Xiao R., Harrison C., Hudson S.~E.}:
\newblock Worldkit: rapid and easy creation of ad-hoc interactive applications
  on everyday surfaces.
\newblock In \emph{Proceedings of the SIGCHI Conference on Human Factors in
  Computing Systems} (2013), ACM, pp.~879--888.

\end{thebibliography}


\end{document}